\documentclass[english,useAMS, usenatbib]{mn2e}
\usepackage{fancyhdr}
\usepackage{float}
\usepackage{subfig}
\usepackage[T1]{fontenc}
\usepackage[latin9]{inputenc}
\usepackage{units}
\usepackage{rotating}
\usepackage{url}
\usepackage{amsmath}
\usepackage{amssymb}
\usepackage{graphicx}
\usepackage{esint}
\usepackage{hyperref}
\usepackage[para,online,flushleft]{threeparttable}

\makeatletter

\def\gtsima{$\; \buildrel > \over \sim \;$}
\def\ltsima{$\; \buildrel < \over \sim \;$}
\def\prosima{$\; \buildrel \propto \over \sim \;$}
\def\gsim{\lower.5ex\hbox{\gtsima}}
\def\lsim{\lower.5ex\hbox{\ltsima}}
\def\simgt{\lower.5ex\hbox{\gtsima}}
\def\simlt{\lower.5ex\hbox{\ltsima}}
\def\simpr{\lower.5ex\hbox{\prosima}}
\def\la{\lsim}
\def\ga{\gsim}

\providecommand{\tabularnewline}{\\}

\title[UVB with metals]{UV background fluctuations traced by metal ions at $z\approx3$}
\author[L. Graziani et al.]{L. Graziani$^{1,2,3}$
\thanks{E-mail: luca.graziani@sns.it}, A. Maselli$^{4}$, U. Maio$^{5}$\\
$^{1}$Max-Planck-Institut f\"ur Astrophysik, Karl-Schwarzschild-Stra{\ss}e 1, D-85748 Garching b. M\"unchen, Germany \\
$^{2}$Scuola Normale Superiore, Piazza dei Cavalieri 7, 56126 Pisa, Italy \\
$^{3}$Dipartimento di Fisica, Sapienza, Universit$\grave{a}$ di Roma, Piazzale Aldo Moro 5, 00185, Roma, Italy\\
$^{4}$Laboratory of Neuromotor Physiology, Santa Lucia Foundation, Rome, Italy\\
$^{5}$Leibniz-Institut f\"ur Astrophysik, An der Sternwarte 16, D-14482 Potsdam, Germany}

\usepackage{babel}
\usepackage{natbib}

\begin{document}

\date{Accepted 2010 <Month> XX. Received 2010 <Month> XX; in original form
2010 <Month> XX}

\maketitle
\pagerange{\pageref{firstpage}--\pageref{lastpage}} \pubyear{2010}\label{dat:firstpage}

\begin{abstract}

Here we investigate how LyC-opaque systems present in the intergalactic medium at $z\approx3$ can distort the spectral shape of a uniform UV background (UVB) through radiative transfer (RT) effects. With this aim in mind, we perform a multi-frequency RT simulation through a cosmic volume of $10h^{-1}$~cMpc scale polluted by metals, and self-consistently derive the ions of all the species. The UVB spatial fluctuations are traced by the ratio of He$\, \rm \scriptstyle II\ $ and H$\, \rm \scriptstyle I\ $ column density, $\eta$, and the ratio of C$\,{\rm {\scriptstyle IV\ }}$ and Si$\,{\rm {\scriptstyle IV\ }}$ optical depths, $\zeta$. We find that: (i) $\eta$ spatially fluctuates through over-dense systems ($\Delta$) with statistically significant deviations $\delta\eta >25$\% in 18\% of the volume ; (ii) same fluctuations in $\zeta$  are also present in $34$\% of the enriched domain (only 8\% of the total volume) and derive from a combination of RT induced effects and in-homogeneous metal enrichment, both effective in systems with $\Delta > 1.5$.

\end{abstract}


\section{INTRODUCTION\label{sec:INTRODUCTION}}

Since the first observations, the He$\,{\rm {\scriptstyle II\ }}$ opacity of the intergalactic medium (IGM) at $z\approx3$ has been interpreted as  ``patchy\textquotedblright{} and highly in-homogeneous \citep{1997A&A...327..890R,1997AJ....113.1495H,2000ApJ...534...69H,2002ApJ...564..542S,
2011ApJ...726..111S,2014ApJ...784...42S}, as highlighted by the Ly-$\alpha$ forest parameter $\eta$\footnote{$\eta$ is defined as He$\, \rm \scriptstyle II\ $ to H$\, \rm \scriptstyle I\ $ column density ratio: $\eta \equiv N_{\textrm{HeII}}/N_{\textrm{HI}}$  \citep{2009RvMP...81.1405M}.}.
In a photo-ionised IGM, $\eta$ is proportional to the UV background spectral shape (UVBSS) and its scatter in space could reflect spatial fluctuations of the UVB at the ionisation edges of H$\,{\rm {\scriptstyle I\ }}$ and He$\,{\rm {\scriptstyle II\ }}$ \citep{1993MNRAS.262..273M}. In the last twenty years different interpretations were provided on the origin of these fluctuations. Spectroscopic observations reported variations of $\eta\in[20-200]$ over scales of $\left[2\text{\textendash}10\right]$ Mpc \citep{1999AJ....118.1450S,1998AJ....115.2206F,2007A&A...461..847F}, which could be interpreted as a local ionisation effect in the proximity of quasars \citep{2004ApJ...600..570S,2010ApJ...722.1312S,2007A&A...473..805W}. A decrease of $\eta$ in redshift, on the other hand, could indicate an evolution in the UVBSS \citep{2004ApJ...605..631Z}. Optical depth ratios from metal lines  have been often  suggested as additional/independent probes of UVBSS spatial variations: $\zeta\equiv \tau_{\textrm{SiIV}}/\tau_{\textrm{CIV}}$ for example, is sensitive to the UVBSS on either side of the He$\,{\rm {\scriptstyle II\ }}$ ionisation edge \citep{1995Natur.375..124S,1996AJ....112..335S,1997AJ....113.1505G,1997A&A...318..347S,1998AJ....115.2184S} and to $[\texttt{Si} / \texttt{C}]$\footnote{$[\texttt{Si} / \texttt{C}] \equiv \log\left(\frac{n_{\texttt{Si}}}{n_{\texttt{C}}}\right)-\log\left(\frac{n_{\texttt{Si}}}{n_{\texttt{C}}}\right)_{\odot}$, $\left(\frac{n_{\texttt{Si}}}{n_{\texttt{C}}}\right)_{\odot} \approx 0.142$. This value depends on the relative abundance $n$ of C and Si polluting the gas and the solar composition model. Here we follow \cite{1998SSRv...85..161G} to be consistent with the  adopted version of Cloudy.}. This optical depth ratio was observed to abruptly change around $z=3$ by \citet{1998AJ....115.2184S} and interpreted as a sudden hardening of the UVB; a redshift evolution of $\zeta$ was also found by \citet{2005A&A...441....9A,2007A&A...461..893A} and \citet{2008A&A...483...19L}. Independent measurements, on the other hand, did not confirm the previous findings \citep{2002A&A...383..747K,2004ApJ...602...38A}. Thus far observations are too scarce to draw a definitive conclusion on the amplitude of the UVBSS fluctuations and on their significance level (e.g. see \citealt{2014MNRAS.440.2406M} for a case against fluctuations). 

Most models of the UVB are not suitable to address this problem because they are generally limited by the assumption of spatial homogeneity and do not account for RT effects, important when density contrasts are present (\citealt{2005MNRAS.364.1429M}, hereafter MF05; \citealt{2006MNRAS.366.1378B,2009ApJ...704L..89M,2012MNRAS.423..558C,2012MNRAS.423....7M,2014MNRAS.437.1141D,2017MNRAS.465.2886D}). Bolton and Viel (2011, hereafter BV11) adopted a set of spatially in-homogeneous UVB models confirming fluctuations in $\eta$ but they did not find any significant difference in the values of $\zeta$ across models. 

By performing a RT simulation of a spatially uniform UVB\footnote{The generated UVB shape and intensity,  as well as its uniform emission in space, are pre-assumed from a selected model with no explicit assumptions on source properties.} through a realistic cosmic web polluted by metals, here we show that both random gas clumps of the IGM  and RT effects can induce spatial fluctuations in the UVBSS. These distortions are traced as scatter in space of $\eta$ and $\zeta$ and computed by the new features of the cosmological RT code \texttt{CRASH3} (\citealt{2013MNRAS.431..722G}, hereafter GR13). We find that: (i) non negligible spatial oscillations of $\eta$ and $\zeta$ come naturally from both RT effects and in-homogeneous distribution in space of gas over-densities $\Delta$\footnote{$\Delta\equiv \rho/\overline{\rho}$, where $\rho$ is the gas density in a point of space (hereafter a cell of a grid) and $\overline{\rho}$ the volume averaged value.}; (ii) $\zeta$ can be used as tracer of UVBSS in the metal enriched sub-domain but one has to be aware of the fact that it suffers the complex interplay between radiative and chemical feedback. 

This paper is structured as follows. In Section 2 we introduce the numerical simulations, while Section 3 discusses the fluctuations in $\eta$ and $\zeta$. Section 4 finally summarizes the conclusions.

\section{Numerical simulations}

The galaxy formation simulation adopted in this work was performed 
on a box of $10h^{-1}$ Mpc comoving by using a modified version of 
the GADGET-2 code \citep{2005MNRAS.364.1105S}, as described  
in \citet{2010MNRAS.407.1003M}. The code includes the whole set 
of chemistry reactions leading to molecule creation and destruction, 
as well as metal production and spreading. Metals are released by AGB stellar winds
and supernovae (SNII, SNIa) explosions following star formation and
according to the stellar lifetimes and production yields. They are successively spread over 
the neighbours of stellar particles according to the SPH
kernel \citep{2007MNRAS.382.1050T} within a kinetic feedback scheme accounting for winds with 
velocity of 500~km~s$^{-1}$. A cosmic UVB (\citealt{1996ApJ...461...20H}, hereafter HM96) is also included as photo-ionisation radiation field. 

One density snapshot at redshift $z\approx3$ was ionized after mapping gas and metals on a Cartesian grid of resolution $N_{c}^{3}=128^3$ cells. Its cosmic web is enriched in $\approx 23$\% of the total domain:
$\approx 17$\%  has metallicity $Z<0.1Z_{\odot}$\footnote{Hereafter the gas metallicity (or equivalently the metal mass fraction) is defined as $Z \equiv M_{Z}/M_{\texttt{gas}}$, where $M_{Z}$ is the total mass
of the elements with atomic number higher than 2, and $M_{\texttt{gas}}$ is the
total mass of the gas. $Z_{\odot} = 0.0126$ as in \citet{2004A&A...417..751A}.}, while only $ \approx 1$\% has $Z > 1Z_{\odot}$. The contribution of metals to the gas cooling function adopted in the RT simulation is also accounted for, as described in GR13.

\begin{figure}
\vspace{-20pt}
\centering
\makebox[0.30\textwidth][c]{
\includegraphics[width=0.5\textwidth,angle=0]{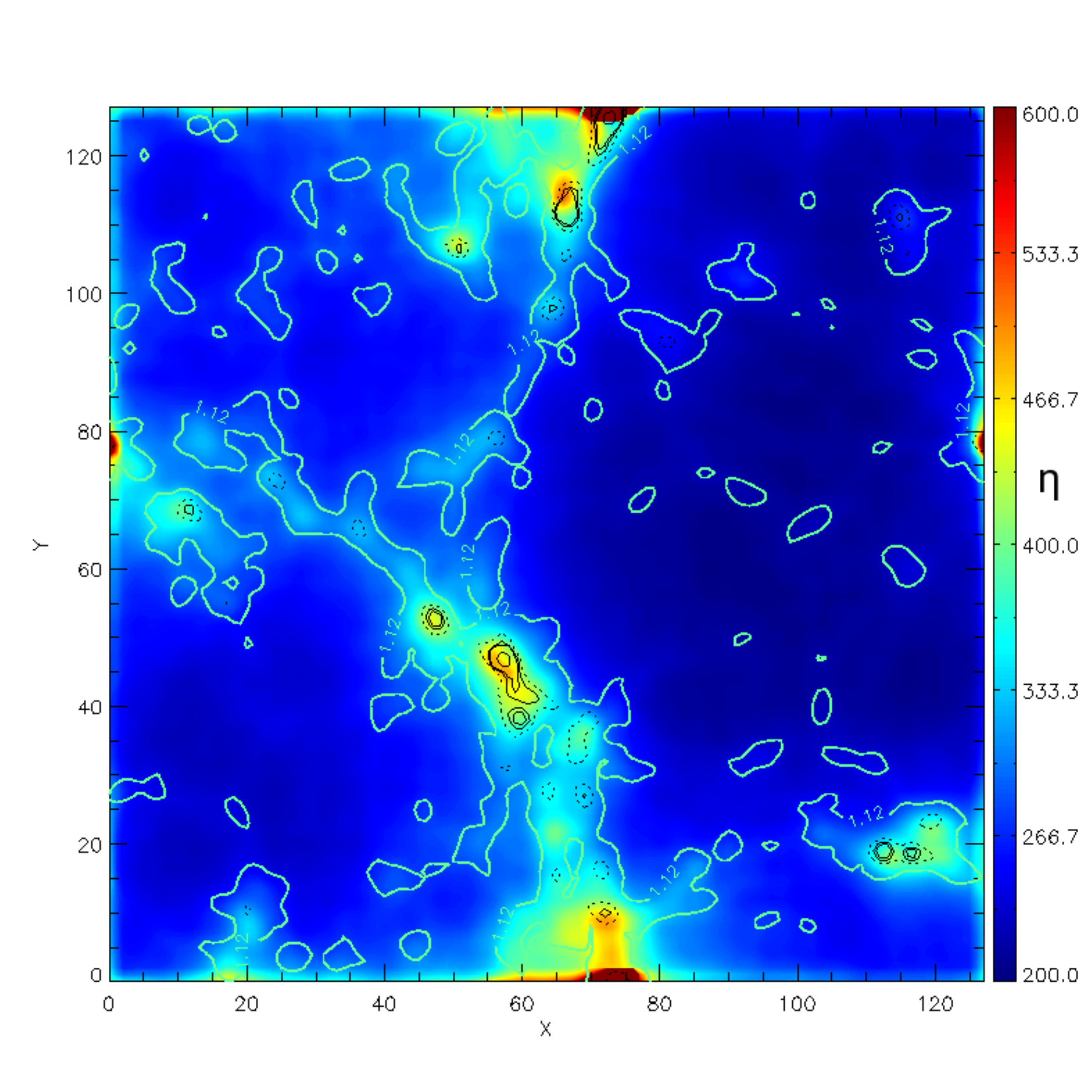}
}
\vspace{-15pt}
\caption{Slice cut through the simulation box showing the spatial distribution of 
         $\eta$ at the equilibrium time $t_{eq}$. The distance in the x- and y-axis is 
         shown in cell units (1 cell $\approx 78$~h$^{-1}$~ckpc), the depth of the slice is 1 cell. 
         Iso-contour lines refer to $\Delta \approx 1$ (white solid), 
         $\Delta \approx 5$ (black dashed), $\Delta \approx 10$ (black solid).}
\label{fig:FigETAPlane}         
\end{figure}

For consistency with the hydrodynamic simulation, \texttt{CRASH3} adopts a HM96 UVB sampled by emitting $10^{4}$ photon packets from $32^3$ grid nodes uniformly covering the cosmic volume\footnote{More details on the UVB numerical scheme can be found in MF05 and will be provided in Graziani et al., in prep. Here note that in our simulation the adoption of a uniform emission grid minimizes by design the intrinsic Monte Carlo noise and guarantees the UVB uniformity assumption with high precision (see Section \ref{sec:EtaResults}).}. The HM96 spectral shape, defined as $J(\nu)/J_{912}$ \footnote{As usual, $J(\nu)$ is the spectrum at frequency $\nu$ and $J_{912}$ is its value at 912~\AA.}, is sampled by $101$ bins, more concentrated around the ionisation frequencies of H and He and extends up to $\nu_{\texttt{max}}\approx 124$~\AA  ($E_{\nu,\texttt{max}}\approx 100$~eV) to allow a direct comparison with the MF05 results\footnote{As \texttt{CRASH3} treats the gas ionization by UV photons and does not account for the physics of X-rays and secondary ionization (but see \citealt{2018MNRAS.479.4320G} for a novel implementation), only the UV range in the original HM96 UV/X-ray spectrum has been selected.}. During the run it is tracked in regions with $Z > 0$ to derive metal ions\footnote{Note that the radiation tracking is  not necessary to compute H and He ionisation (see GR13 for more details).} with the \texttt{CLOUDY} (v.10, \citealt{1998PASP..110..761F}) module embedded in \texttt{CRASH3}. Periodic boundary conditions are set up by reusing the escaping packets 10 times to approximately cover the mean free path of ionising photons at $z = 3$ ($\approx 100 h^{-1}$~cMpc, \citealt{2002AJ....123.1247F}). The simulation starts from a neutral gas at $T_0=100$ K and proceeds until an ionisation equilibrium is reached ($t_{eq} = 5.5\cdot10^{6}$~yr)\footnote{Initial conditions consistent with the reionization history could be obtained from reionisation simulations of both H and He. We defer this investigation to a future work having a consistent calculation from reionization simulations accounting for metals.}. The resulting volume averaged ionisation fractions are $\overline{x}_{\texttt{HII}} \approx 0.999997$, $\overline{x}_{\textrm{HeII}}\approx 0.036899$ and $\overline{x}_{\textrm{HeIII}}\approx0.963097$ \footnote{Note that the total number of photons crossing the domain is higher than $10^9$ and guarantees Monte Carlo convergence up to the $10^{-6}$ in the hydrogen ionisation fraction, at equilibrium.}

\section{Results \label{sec:Results} }

Here we show the spatial fluctuations of $\eta$ and $\zeta$ at $t_{eq}$ and we study them as function of $\Delta$. The relative fluctuation of $\eta$ (or $\zeta$) around the highest probability value ($\eta_{p}$) of its distribution\footnote{Operationally, the highest probability value is the center of the histogram bin containing the largest number of cells.} is defined as $\delta \eta \equiv \left| \dfrac{\eta-\eta_{p}}{\eta_p}\right|$, while $\overline{\eta}$ indicates the volume averaged value.
The statistical distributions of both quantities are computed in all the cells in which they are investigated, i.e. the cosmic volume for $\eta$ and the over-dense, metal polluted domain for $\zeta$.

\subsection{Fluctuations of $\eta$ \label{sec:EtaResults}}

To compute $\eta$ at simulation run-time we adopt the notation of \citet{1998AJ....115.2206F}:

\begin{equation}
\eta = \frac{\alpha_{\textrm{HeII}}(T)}{\alpha_{\textrm{HI}}(T)}\frac{n_{\textrm{HeIII}}}{n_{\textrm{HII}}}\frac{\Gamma_{\textrm{HI}}}{\Gamma_{\textrm{HeII}}},\label{eq:EthaNumeric}
\end{equation}

where $\alpha_{\textrm{HeII}}$ and $\alpha_{\textrm{HI}}$ are the Case A He$\,{\rm {\scriptstyle II\ }}$ and H$\,{\rm {\scriptstyle I\ }}$ recombination coefficients, while $\Gamma_{\textrm{HeII}}$ and $\Gamma_{\textrm{HI}}$ are their photo-ionisation rates computed in each cell of the domain.

In Figure \ref{fig:FigETAPlane} we show $\eta$ in a slice with a central over-dense filament enriched by metals. The under-dense regions ($\approx33$\% of the plane, dark-blue areas) are characterized by $\eta \approx 260$, while the central filament shows  values in the range $350<\eta<450$. Higher values, up to $\eta = 600$, are found instead in the densest clumps visible at the borders of the figure\footnote{Note that the periodic boundary conditions applied to the hydrodynamical simulation create quasi-symmetric over-density areas at the volume edges.}. To highlight the tight correlation between $\eta$ and $\Delta$, we over-plot $\Delta$ iso-contour lines. As argued in MF05, at photo-ionisation equilibrium $\eta$ increases with $\Delta$ for a combination of reasons: (i) the recombination rates show a weak dependence on $T$ and then the ratio is expected to weakly reflect the increase of $T$ in $\Delta > 1$ \citep{1998MNRAS.301..478T}; (ii) helium recombines $5.5$ times faster than hydrogen and in over-dense regions the ratio $\frac{n_{\textrm{HeII}}}{n_{\textrm{HI}}}$ drives the increase of $\eta$ with $\Delta$ as the Universe is more opaque in He$\,{\rm {\scriptstyle II\ }}$; (iii) $\Gamma_{\textrm{HI}}$ is always one order magnitude larger than $\Gamma_{\textrm{HeII}}$. Note that spatial fluctuations in $T$ are also present and will be investigated in a companion paper.

\begin{figure}
\vspace{-1.5pt}
\centering
\makebox[0.30\textwidth][c]{
\includegraphics[scale=0.28,angle=0]{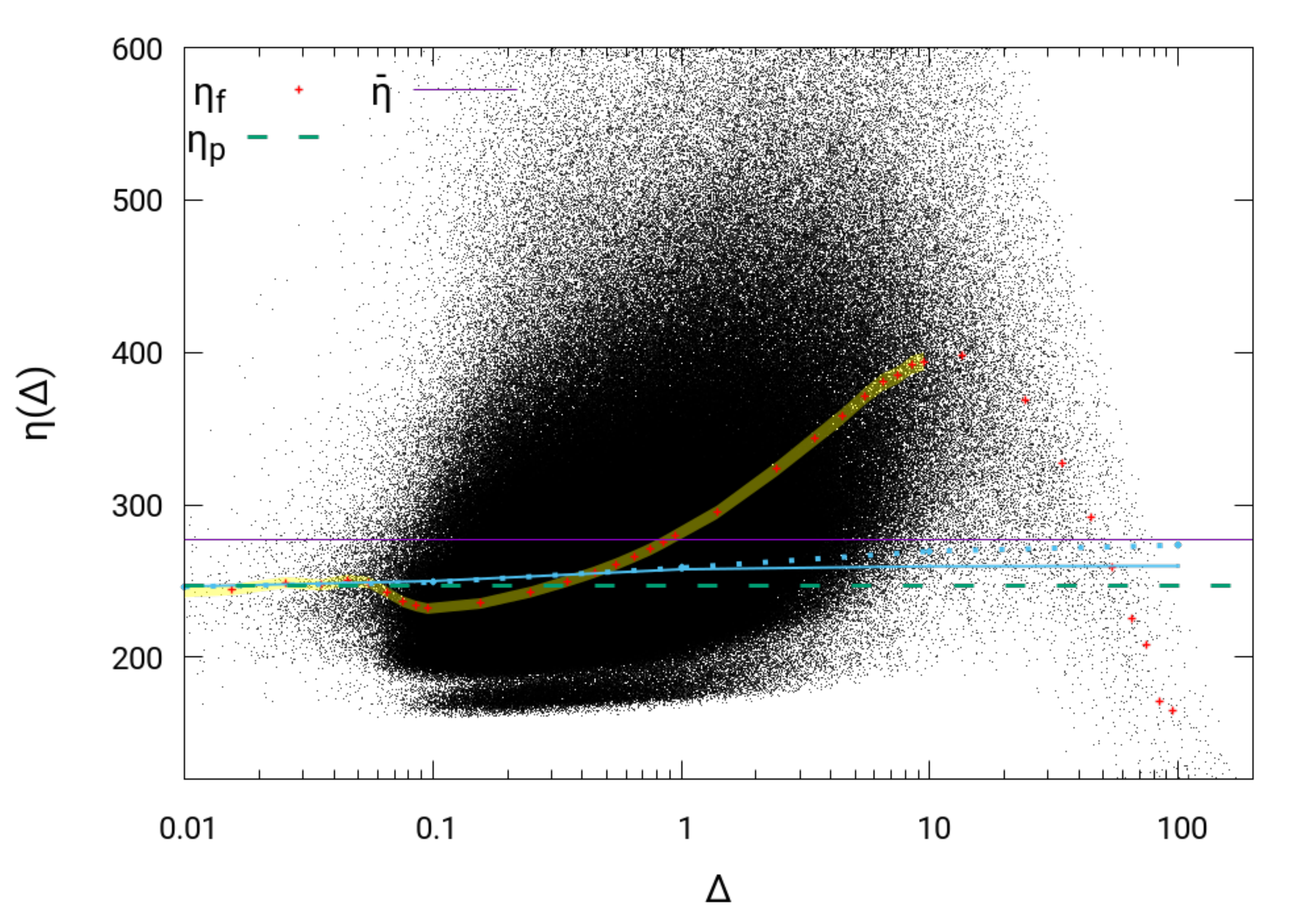}
}
\caption{Scatter plot of $\eta$ (black points) as function of $\Delta$ at the equilibrium time $t_{eq}$. The average value at fixed $\Delta$, $\eta_f$, is shown as red crosses, while the highest probability value $\eta_p$ in the volume as dashed green line and the volume average value  $\overline{\eta}$ as violet solid line. The yellow shadow area indicates the average noise in $\eta_f$ intrinsic in our RT algorithm. Finally, dotted (solid) light-blue lines show the values of $\eta_{\texttt{eq}}(\Delta, Z)$ assuming a uniform HM96 UVB and gas metallicity $Z = 0.01$~Z$_\odot$ ($Z=3$~Z$_\odot$).
}
\label{fig:FigETA}
\end{figure}

Figure \ref{fig:FigETA} shows, as scatter plot, the values of $\eta$ found in each cell of the cosmic volume as function of their over-density $\Delta$. The average of $\eta$ at fixed over-density (${\eta_f}$) is also shown as reference with red crosses. The yellow transparent area indicates the effects of the \texttt{CRASH3} numerical noise  on $\eta$, evaluated by propagating a spatially homogeneous UVB through a medium with constant number density \footnote{Here we checked the cases corresponding to $n_{gas} \sim 10^{-7}, 10^{-6}, 10^{-5}$ (i.e. $\Delta \approx 1$)$,10^{-4}$~cm$^{-3}$. This tests minimize the RT effects and allow to establish the method intrinsic noise once the Monte Carlo convergence is established (see Section 2).}; the fluctuations induced on $\eta$ by that noise are always lower than 1.5\% in all tests. A significant scatter is found around each ${\eta_f}$ (red crosses) and around both $\eta_p$ (violet solid line) and $\overline{\eta}$ (dashed green line), computed from the global statistic.  Note that regions  with $\Delta<0.1$ exhibit a low scatter at fixed number density: 5-8\% around ${\eta_f} \approx 252$, with only few cells reaching 15\%. In $0.1<\Delta \leq 10$, $\eta_f$ increases with $\Delta$ and the scatter shows typical statistically significant  variations of 20-30\%, with values as high as 60\% in the range $7 \leq \Delta \leq 10$\footnote{While the scatter plot shows points largely deviating from the average values in the entire domain, their significance is established by looking at their statistical distribution in fixed over-density bins.}. As discussed in MF05, in very dense regions with $\Delta>10$ recombination dominates over ionisation, inducing an inversion of the trend, with $\eta_f$ decreasing with increasing $\Delta$. Finally note that the small cloud of points in which $\eta < 200$ ($0.1 < \Delta < 10$) is created by cells hosting photon emission nodes. In these cells in fact our ionization algorithm tends to fully ionize the gas and to lower the value of $\eta$; their total number is, on the other hand, lower than 1\% with a negligible impact on the global statistic. As a comparison, we computed $\eta_{\texttt{eq}}(\Delta, Z)$, i.e. the value of $\eta$ determined by \texttt{Cloudy} photo-ionisation equilibrium models assuming a spatially homogeneous HM96 UVB, as described in Section 2\footnote{Formula \ref{eq:EthaNumeric} shows that $\eta_{\texttt{eq}}$ can vary because of both $\alpha_{\textrm{HeII}}(T)/\alpha_{\textrm{HI}}(T)$ and $n_{\textrm{HeIII}}/n_{\textrm{HII}}$, at assigned UVB and $E_{\nu,\texttt{max}}$. These terms depend on $Z$ and $n_{\texttt{gas}}$ in each cell; the total computation can then be performed by a grid of models spanning their values. Finally note that the choice of $E_{\nu,\texttt{max}}$ implies a less steep $\eta_{\texttt{eq}}(\Delta)$ relation with respect to adopting the original HM96 spectral range as the X-ray contribution to  helium ionisation is not accounted for \citep{2018MNRAS.479.4320G}.}. All $\eta_{\texttt{eq}}(\Delta,Z)$ lie between the values corresponding to the min/max of the gas metallicity ($0.01$~Z$_\odot$, dotted; $3$~Z$_\odot$, solid light-blue lines). In under-dense regions $\eta_{\texttt{eq}} \approx \eta_f$ because the RT effects are minimized and $Z$ is usually low. $\eta_{\texttt{eq}}$ slightly increases with $\Delta$ at all $Z$ remaining confined in $\eta_p < \eta_{\texttt{eq}}(\Delta) < \overline{\eta}$; the divergence between dotted and solid light-blue lines is caused by the decrease of $T$ at increasing $Z$ but with modest effects because $\alpha_{\textrm{HeII}}/\alpha_{\textrm{HI}} \propto (T/10^{4.3})^{0.06}$. Finally note that at $\Delta > 1$ the RT effects become important and then $\eta_{\texttt{eq}}$ rapidly separate from $\eta_f$.

The statistical distribution of $\eta$  shows $\eta_p \approx 247$ and $\overline{\eta} \approx 277$. In $\approx 82$\% of the domain $\delta\eta \leq 25$\%, while a fluctuation within $25$\%$ < \delta\eta \leq 50$\% is found in 11\% of the cells. Finally, values in $50$\%$ < \delta\eta \leq 75$\% are found in 4\% of the domain and a remaining 3\% shows $\delta\eta > 75$\%.

The above results are in global agreement with MF05, but it should be noted that our statistic shows higher fluctuations when $\Delta > 1$ due to the improved feedback model of the hydrodynamical simulation, which creates sharper density gradients and enhances the RT effects. As a result, the spatial UVB fluctuations increase. Finally, we point out that while our analysis is not meant to quantitatively reproduce observations (this will be addressed in a companion project with an updated UVB model, i.e. \citealt{2012ApJ...746..125H}), the predicted value of $\overline{\eta} \approx 277$ is consistent with the one observed in \citet{2000ApJ...534...69H} and \citet{2014ApJ...784...42S} at $z \approx 3$. Our fluctuation range is reasonable as well, as it is consistent with the estimates in \citet{2004ApJ...600..570S}  inferred at a similar scale (i.e $10h^{-1}$ Mpc comoving), although it is in tension with values found in \citet{2014MNRAS.440.2406M}. Note, though, that the latter results refer to lower redshift data ($z<2.7$), where the progress of reionization could have significantly reduced the fluctuations of $\eta$.
  
\subsection{Fluctuations of $\zeta$ \label{sub:ZETA}}

Here we discuss the scatter in $\zeta$, which provides an additional evidence of the UVBSS spatial fluctuations using metal ions. Since \texttt{CRASH3} does not include the metal contribution to the gas optical depth (see GR13 for more details), $\zeta$ computed in this paper fluctuates as a result of absorption by H and He only, and of in-homogeneous chemical enrichment. Spectral distortions around the ionisation energy of H$\,{\rm {\scriptstyle I\ }}$ and He$\,{\rm {\scriptstyle II\ }}$ impact in fact the ionization of both Si$\,{\rm {\scriptstyle IV\ }}$ and C$\,{\rm {\scriptstyle IV\ }}$ and alter the ratio of their ionization fractions. In addition, fluctuations of $[\texttt{Si} / \texttt{C}]$ are expected in different density environments because metals are followed individually during their spreading outside the  formation sites. For an easier comparison with BV11, we compute $\zeta$ in each cell of the over-dense, polluted volume as: 

\begin{equation}
	\zeta = 1.7 \frac{x_{\texttt{SiIV}}}{x_{\texttt{CIV}}}10^{[\texttt{Si} / \texttt{C}]-0.77},
\end{equation}  

where $x_{\texttt{SiIV}}$ and $x_{\texttt{CIV}}$ are the ionization fractions of Si$\,{\rm {\scriptstyle IV\ }}$ and C$\,{\rm {\scriptstyle IV\ }}$, respectively. $\zeta$ can then be studied as a combination of two terms: $x \equiv x_{\texttt{SiIV}} / x_{\texttt{CIV}}$ (mainly affected by RT) and $y \equiv 10^{[\texttt{Si} / \texttt{C}]-0.77}$, reflecting the in-balance of atom abundances created by mechanical and chemical feedback. 
Hereafter we focus on the sub-domain $\Delta > 1.5$ because it is significantly metal polluted and simultaneously shows large fluctuations in the UVB spectral shape through $\eta$.

\begin{figure}
\makebox[0.45\textwidth][c]{
\includegraphics[scale=0.25,angle=0]{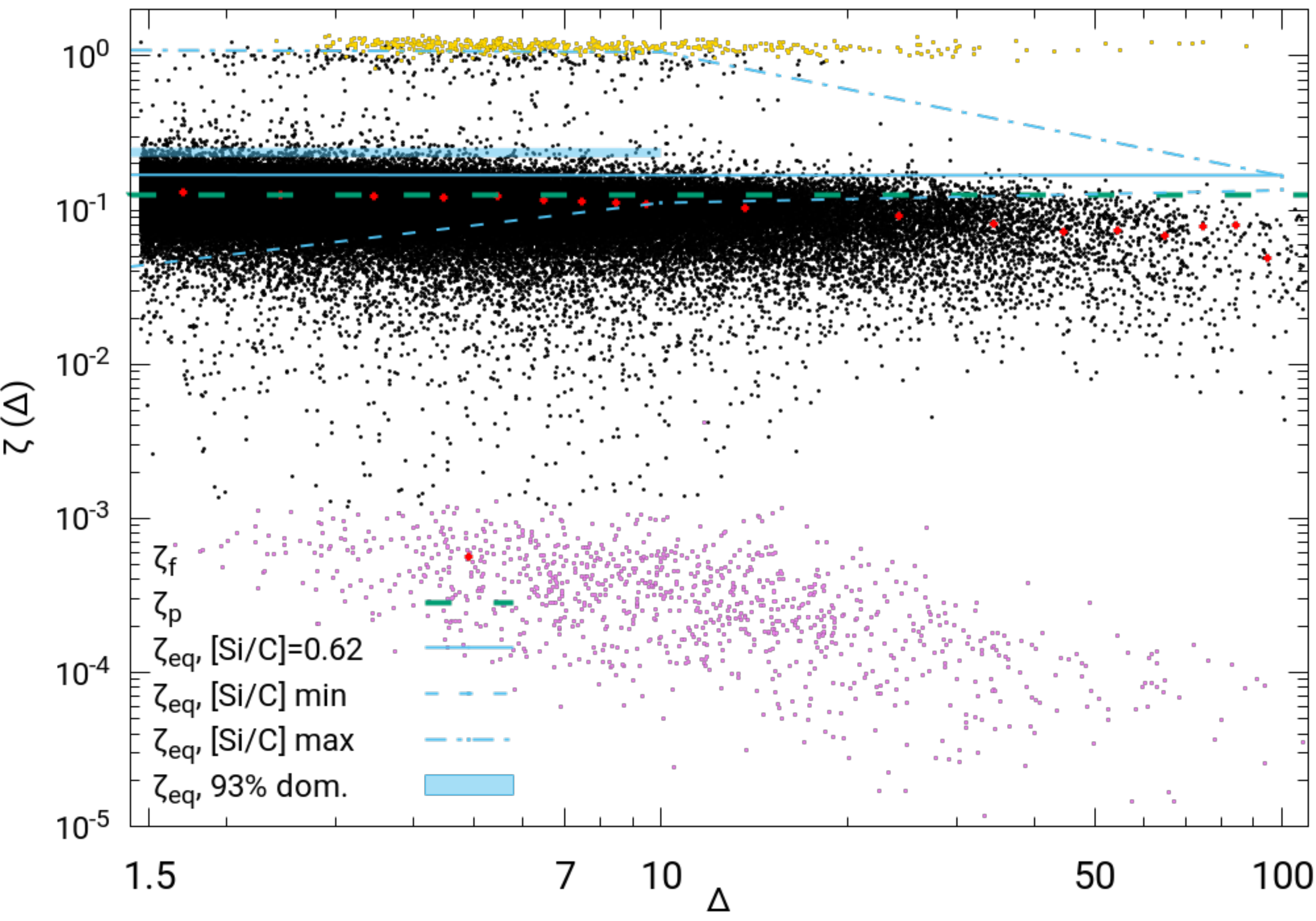}
}
\vspace{-1.5pt}
\caption{Scatter plot of $\zeta$ (black points) as function of $\Delta$ at $t_{eq}$. The average value at fixed $\Delta$, $\zeta_f$, is shown as red crosses, while the highest probability value $\zeta_p$ as dashed green line. Violet points indicate cells in which $x \equiv x_{\texttt{SiIV}} / x_{\texttt{CIV}} < 10^{-3}$, while gold points mark cells in which $x = 1$. Dashed/dashed-dotted light-blue lines show $\zeta_{\texttt{eq}}(\Delta, [\texttt{Si} / \texttt{C}])$ computed at the minimum/maximum values of $[\texttt{Si} / \texttt{C}](\Delta)$, while a case at the highest probability value of $[\texttt{Si} / \texttt{C}](1.5)\approx 0.62$ is shown as a solid line. Finally note that in about 93\% of the domain $1.5 \leq \Delta < 10$, $\zeta_{\texttt{eq}}$ has values in the cyan shadow area.}
\label{fig:FigZETA}
\end{figure}

Figure \ref{fig:FigZETA} shows the scatter plot of $\zeta(\Delta)$. Deviations of $\zeta$ from $\zeta_f$ are present everywhere, but contrary to $\eta$ there is no tight dependence on $\Delta$ and $\zeta_f$ decreases of about 25\% from $\Delta < 7$ ($\zeta_f\approx 0.125$) to higher over-densities where $\zeta_f\approx 0.1$. 
Large fluctuations (up to several orders of magnitude) are clearly visible around any $\zeta_f$, but when a statistically significant number of cells is considered their amplitude reduces to $\approx$60\%. In specific cases (see the cloud of violet points) $\zeta$ scatters in $10^{-5} < \zeta < 10^{-3}$ because of the extremely low values of $x$ ($x < 10^{-3}$), i.e. either the spectral shape created in these cells is highly altered by absorption at the ionization potential of Si$\,{\rm {\scriptstyle IV\ }}$ ($\approx 45$~eV) or a hard spectrum generates higher ionization states (see for example the model UVB3 studied in BV11). Another limited set of points (gold cloud) is determined by the conditions: $x = 1$, and $y$ almost constant. In these cells either Si$\,{\rm {\scriptstyle IV\ }}$ and C$\,{\rm {\scriptstyle IV\ }}$ are fully ionized or in ionization states higher than IV.
As in Figure \ref{fig:FigETA}, the light-blue lines refer to $\zeta_{\texttt{eq}}(\Delta, [\texttt{Si} / \texttt{C}])$ computed by assuming a uniform UVB. At constant  $[\texttt{Si} / \texttt{C}](1.5)\approx 0.62$ (i.e. $y(\Delta)\approx 0.7$, solid line), $\zeta_{\texttt{eq}}(\Delta)$ shows a global decrease of only 4\%, driven by a decrease of $x$ with increasing $\Delta$. When the UVB is assumed homogeneous in space, the only term inducing significant spatial fluctuations is then $y$. By adopting the min/max values of $y(\Delta)$ we can trace the dashed/dashed-dotted lines, also showing that the maximum variation of $[\texttt{Si} / \texttt{C}]$ is found in $1.5 < \Delta < 10$, while it significantly reduces at higher over-density. These extreme cases, on the other hand, must be taken with a grain of salt as they are found in less than 7\% of the cells; the scatter of $[\texttt{Si} / \texttt{C}]$ from the average value is $\leq 4$\% in the remaining 93\%, i.e. $\zeta_{\texttt{eq}}$ is confined in the cyan shadow area.\footnote{Note that (i) the hydrodynamical scheme shows an average $[\texttt{Si} / \texttt{C}]\approx 0.77$, perfectly consistent with the one observed in the IGM \citep{2004ApJ...602...38A}; (ii) the trend of $\zeta_{\texttt{eq}}(\Delta)$ strictly depends on the specific $E_{\nu, max}$ adopted in the computation, as the metal ionization reacts to a spectral range wider than the one of H and He (see GR13 and \citealt{2018MNRAS.479.4320G} for more details).}.

\begin{figure}
\vspace{-1.5pt}
\centering
\makebox[0.30\textwidth][c]{
\includegraphics[scale=0.28,angle=0]{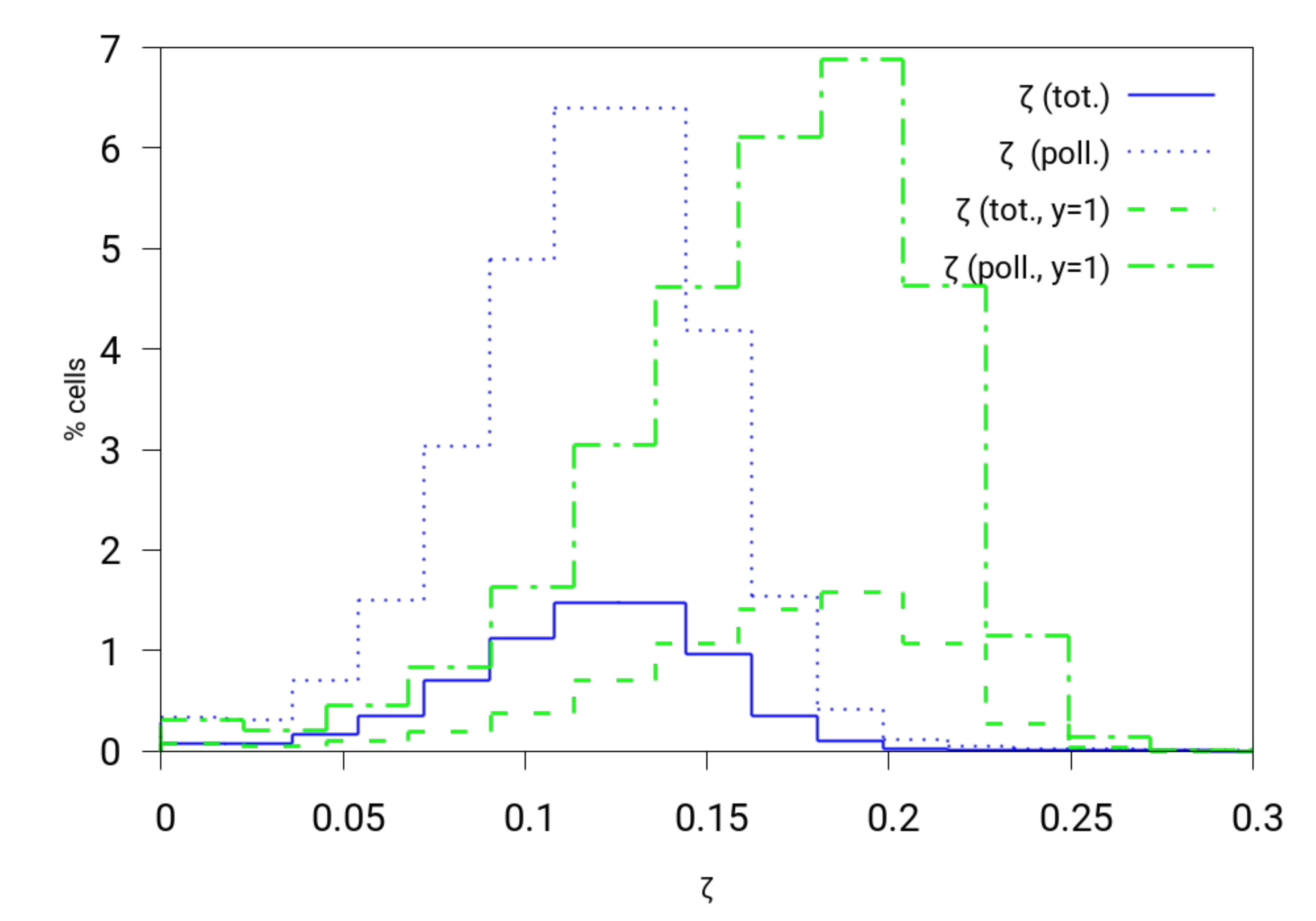}
}
\caption{Percentage of cells with $\zeta$ in $\Delta > 1.5$ at $t_{eq}$ (blue lines). Same case with $y=1$ is shown as green lines. The percentage is determined over the total volume (solid/dashed lines) and the polluted volume (dotted/dashed-dotted).}
\label{fig:statZETA}
\end{figure}

In Figure \ref{fig:statZETA} we show the statistic of $\zeta$ computed over $N_{c}^{3}$ cells (total domain, solid blue lines) and over the total number of cells polluted by metals (polluted domain, dashed blue lines); in this way the reader can have a feeling of the percentage of the cosmic (or enriched) volume in which $\zeta$ can be used as tracer of the UVBSS\footnote{Note that both estimates depend on the enrichment scheme of the simulation.}. To separate the contributions of $x$ and $y$, we also show the analogous histograms obtained from a simulation having $[\texttt{Si} / \texttt{C}] = 0.77$ (i.e. $y=1$, green lines). 

The fluctuations described in Figure \ref{fig:FigZETA} are confirmed by the width of the blue histograms and can now be quantified. Fluctuations of $\delta\zeta \leq 25$\% around $\zeta_p\approx 0.125$ are present in $66$\% of the domain; $25$\%$< \delta\zeta \leq 50$\% in 25\% of cells, $50$\% $ < \delta\zeta \leq 75$\% in 6\% and finally, only 3\% of the cells experience $\delta\zeta > 75$\%. By comparing with the green histograms, the effects of a pure RT feedback come to light: they generate a more asymmetrical distribution, with a peak shifted to $\zeta_p \approx 0.183$. The cells with $\delta\zeta > 25$\% are also systematically reduced because the above domains change to 74\%, 20\%, 4\%, 2\%. We can then conclude that the combined radiative and chemical feedback terms increase the domain in which $\zeta$ shows relevant fluctuations, although the radiative effects remain dominant.

In conclusion, we emphasize that the possibility of using $\zeta$ as additional tracer of UVBSS fluctuations is severely limited by the number of mildly over-dense and polluted systems found in the cosmic web. 

\section{Conclusions \label{sub:UVBShape}}

In this paper we post-process a hydrodynamic simulation which includes metal pollution with the multi-frequency cosmological radiative transfer (RT) code \texttt{CRASH3} to study the amplitude and statistical relevance of spatial fluctuations of the UV background spectral shape (UVBSS) at the epoch of helium reionisation ($z \approx 3$). As the slope of the UVBSS can not be inferred by direct observations, its fluctuations must be constrained by combining the observed scatter in two quantities sensitive to the shape around the He$\,{\rm {\scriptstyle II\ }}$ ionisation potential: $\eta \equiv N_{\textrm{HeII}}/N_{\textrm{HI}}$ and $\zeta \equiv \tau_{\textrm{SiIV}}/\tau_{\textrm{CIV}}$. 
Note that a theoretical investigation of this problem can not be effectively performed with conventional UVB models which do not include an accurate RT. In this work, for the first time in the literature, we employ a radiative transfer approach through H, He and metal species to evaluate $\eta$ and $\zeta$ self-consistently, guaranteeing that the significant spatial fluctuations obtained are indeed due to the combined effect of metal enrichment and radiation transfer. 
In particular:
\begin{itemize}

\item we find a tight correlation of the $\eta$ parameter and over-dense systems of the cosmic web on a
      scale of $10h^{-1}$~cMpc and a resolution of $\approx 78$~h$^{-1}$~ckpc; these spatial fluctuations can reach values         higher than $25$\% in 18\% of the domain and are due to RT effects through the cosmic web; 
  
\item by computing metal ions self-consistently with the RT it is possible to reproduce spatial fluctuations of $\zeta$ higher than $25$\% in 34\% of the metal enriched, over-dense systems with $\Delta > 1.5$ (i.e. $8$\% of the total volume). To be effectively used as tracer of the UVBSS, $\zeta$ requires then the presence of a statistically relevant number of polluted, over-dense systems along observed lines of sight;
      
\item although radiative effects remain dominant, $\zeta$ depends on both UVBSS distortions and spatial fluctuations of $[\texttt{Si} / \texttt{C}]$; we have shown that their combined effects increase the domain in which $\delta\zeta >25$\%. 

\end{itemize}            
Future studies will focus on complementary sources of UVB fluctuations, mainly associated with the variability of quasars at the epoch of helium reionisation, and will compare their statistical significance with the present findings.          

\section*{ACKNOWLEDGMENTS }
The authors are indebted to the anonymous referee for the significant help in improving the paper. We are also grateful to J. Bolton, R. Dav\'{e}, K. Finlator, A. Ferrara, G. Worseck and M. McQuinn for enlightening discussions. LG warmly thanks B. Ciardi and K. Dolag for their contribution to the initial manuscript. AM and LG acknowledge the support of the German Research Fundation (DFG) Priority Program 1177 and 1573, and UM of the DFG project n. 390015701 and the HPC-Europa3 Transnational Access program, grant n. HPC17ERW30.

\bibliographystyle{mn2e}
\bibliography{UVBz3Letter}{}

\label{lastpage}
\end{document}